\documentclass[12pt]{article}
\usepackage[pdftex]{color}
\usepackage{amsmath,amsthm,amssymb}
\usepackage{wasysym}
\usepackage{graphicx}
\usepackage{color}%
\usepackage[onehalfspacing]{setspace}
\usepackage{lipsum}
\usepackage[lmargin=2.5cm,rmargin=2.5cm,tmargin=3cm,bmargin=3cm]{geometry}
\usepackage{multirow}
\usepackage{hyperref}
\hypersetup{hidelinks}
\usepackage{mathtools}
\usepackage{breqn}
\usepackage{threeparttable}
\usepackage[utf8]{inputenc}
\usepackage{tocloft}
\usepackage{cite}
\usepackage{enumitem}

\setlist[itemize]{leftmargin=*}

\begin{document}

%\linenumbers
\begin{center}
{\rm \bf \Large{Stochastic process model for interfacial gap of purely normal elastic rough surface contact}
}
\end{center}

\begin{center}
{\bf Yang Xu$^{\text{ab}}$\footnote{Corresponding author: yang.xu@hfut.edu.cn}, Junki Joe$^{\text{c}}$, Xiaobao Li$^{\text{d}}$, Yunong Zhou$^{\text{e}}$\footnote{Corresponding author: yunong.zhou@yzu.edu.cn}}
\end{center}

\begin{flushleft}
{
$^{\text{a}}$School of Mechanical Engineering, Hefei University of Technology, Hefei, 230009, China\\
$^{\text{b}}$Anhui Province Key Laboratory of Digital Design and Manufacturing, Hefei, 230009, China\\
$^{\text{c}}$1001 Tiverton Ave, Los Angeles, CA 90024, USA\\
$^{\text{d}}$School of Civil Engineering, Hefei University of Technology, Hefei, 230009, China\\
$^{\text{e}}$Department of Civil Engineering, Yangzhou University, Yangzhou, 225127, China \\
}
\end{flushleft}

\begin{abstract}
In purely normal elastic rough surface contact problems, Persson's theory of contact shows that the evolution of the probability density function (PDF) of contact pressure with the magnification is governed by a diffusion equation. However, there is no partial differential equation describing the evolution of the PDF of the interfacial gap. In this study, we derive a convection--diffusion equation in terms of the PDF of the interfacial gap based on stochastic process theory, as well as the initial and boundary conditions. A finite difference method is developed to numerically solve the partial differential equation. The predicted PDF of the interfacial gap agrees well with that by Green's Function Molecular Dynamics (GFMD) and other variants of Persson's theory of contact at high load ranges. At low load ranges, the obvious deviation between the present work and GFMD is attributed to the overestimated mean interfacial gap and oversimplified magnification-dependent diffusion coefficient used in the present model. As one of its direct application, we show that the present work can effectively solve the adhesive contact problem under the DMT limit. The current study provides an alternative methodology for determining the PDF of the interfacial gap and a unified framework for solving the complementary problem of random contact pressure and random interfacial gap based on stochastic process theory. 
\end{abstract}

{\bf Keywords}: Persson's theory; rough surface contact; interfacial gap; purely normal contact; elasticity; stochastic process theory.

\section{Introduction}
Since its introduction more than $20$ years ago \cite{Persson01}, Persson's theory of contact has become one of the dominant approaches for solving rough surface contact problems. Persson's theory is a broad concept including a series of probability-based rough surface contact studies, which can be  categorized as solving either interaction properties (e.g., the probability density function (PDF) of contact pressure \cite{Persson01,Dapp14,Wang17,xu2024persson}, relative contact area \cite{Persson01,Manners06,Yang08,Wang17,Xu24Fudge}, and linear strain energy \cite{Persson07,Persson08,Xu24Fudge}) or gap properties (e.g., the PDF of an interfacial gap \cite{Yang08,Almqvist11,Afferrante18}, and average interfacial gap \cite{Persson07,Persson08}). During its historical development, Persson's theory has been successfully applied to classic tribological problems such as electrical contact \cite{Persson22}, seals \cite{Bottiglione09, Persson16}, electroadhesion \cite{Persson18, Ciavarella20}, and contact electrification \cite{Persson20}.

Let us consider a random, self-affine, isotropic, bandwidth-limited rough surface with the magnification $\zeta$ defined as the ratio of upper cut-off wavenumber to lower cut-off wavenumber. As $\zeta$ increases, random roughness components with smaller wavelengths are superposed onto the existing rough surface topography. Assuming an elastic half-space with a nominally flat rough surface as the boundary in purely normal contact with a rigid flat, Persson \cite{Persson01} derived that the evolution of the PDF of random contact pressure with magnification can be described by a diffusion equation, whose solution is composed of a Gaussian distribution, from which its mirror image about zero pressure is subtracted \cite{Manners06, Yang06}. Several fudge factors \cite{Yang08,Wang17,Xu24Fudge} have been proposed to correct the overestimated diffusion coefficient. Xu et al. \cite{xu2024persson} recently gave an alternative derivation for the diffusion equation based on stochastic process theory and showed that the derivation of the diffusion equation relies on two fundamental assumptions: (1) the variation of random contact pressure with the magnification is a \emph{Markov} process \cite{Risken89, Gillespie92}; and (2) a \emph{no re-entry} assumption \cite{Dapp14} (see Section \ref{sec:PDF_gap} for the detailed explanation). 

The variation of the PDF of the interfacial gap with the magnification receives little attention. In the first attempt, Almqvist et al. \cite{Almqvist11} formulated the PDF of the interfacial gap  in an integral form, which was recently improved by Afferrante et al. \cite{Afferrante18}. Both solutions   show good agreement with various numerical solutions \cite{Almqvist11, Afferrante18}. Recently, Joe et al. \cite{Joe17, Joe18} used a compound Chapman--Kolmogorov equation to solve the PDF of the interfacial gap in adhesive contact problems. This model was further extended by Joe et al. \cite{Joe22, Joe23} to solve the non-adhesive contact by using the power-law type of regularized non-adhesive surface interaction law.

It is commonly known that the interfacial gap and the contact pressure form a linear complementary problem, which satisfies the Karush--Kuhn--Tucker condition \cite{Polonsky99,bemporad2015optimization}. We may expect that the evolution of the random interfacial gap with the magnification is also a Markov process. Thus, a similar partial differential equation may be derived following the   approach of Xu et al. \cite{xu2024persson}. Following this idea, the present study is organized as follows. Section \ref{sec:Persson_theory} briefly introduces the PDF of the contact pressure solved by   Persson's theory of contact. A convection--diffusion equation of the PDF of the interfacial gap is derived based on stochastic process theory in Section \ref{sec:PDF_gap}. The finite difference method for solving the partial differential equation is developed in Section \ref{sec:FDM}. Some representative results and discussion are given in Section \ref{sec:RandD}. 

\section{PDF of contact pressure}\label{sec:Persson_theory}

\begin{figure}[h!]
  \centering
  \includegraphics[width=16cm]{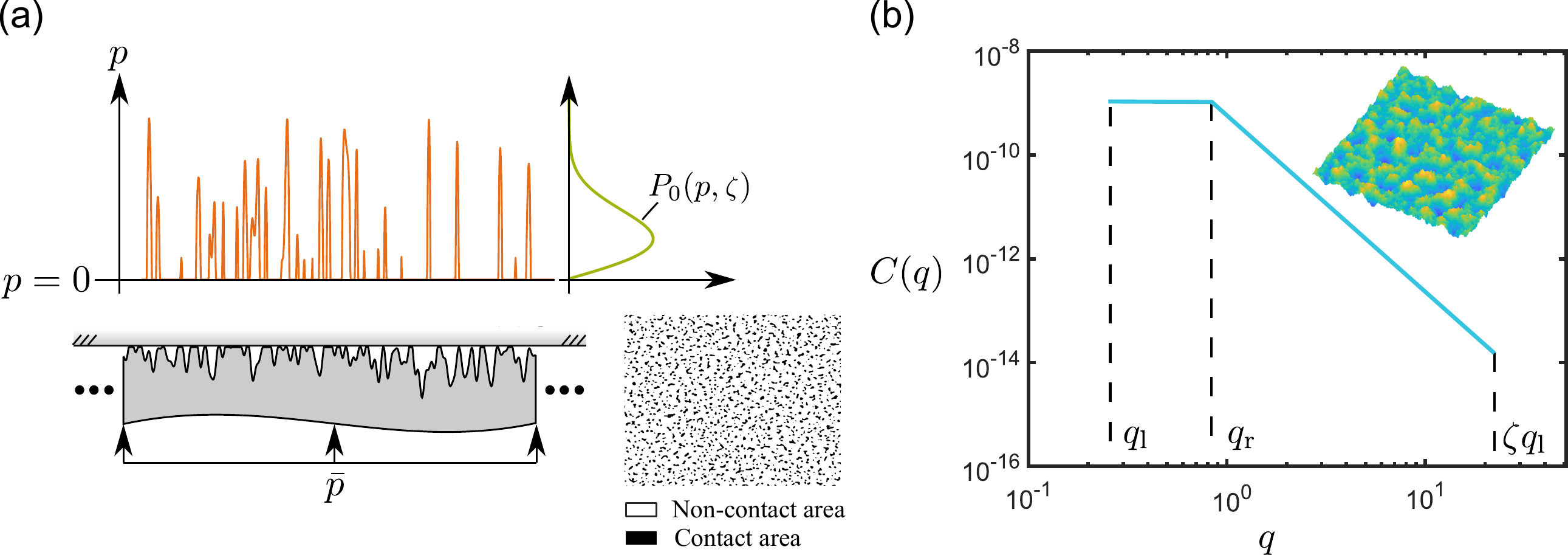}
  %\internallinenumbers 
  \caption{(a) Schematic of cross-section of deformed interface and contact pressure distribution between an elastic half-space with rough boundary and a rigid flat, distribution of contact spots, and PDF of contact pressure $P_0(p, \zeta)$; (b) PSD of a typical random, self-affine, isotropic, bandwidth-limited rough surface, where inset shows one rough surface realization.}\label{fig:Fig_1}
\end{figure}

Let us consider the purely normal contact between an elastic half-space with a nominally flat rough boundary and a rigid flat, see Fig. \ref{fig:Fig_1}(a). The rigid flat is spatially fixed, while the far end of the half-space is subjected to a uniform compressive normal traction $\bar{p}$. The power spectral density (PSD) of the rough boundary has the following piecewise form (see Fig. \ref{fig:Fig_1}(b) for a graphical illustration): 
%\begin{linenomath*}
\begin{equation}\label{E:PSD}
  C(q) =
  \begin{cases}
  C_0 q_{\text{r}}^{-2 (1 + H)} ~~~ q \in [q_{\text{l}}, \min(q_{\text{r}}, \zeta q_{\text{l}}) ), \\
  C_0 q^{-2 (1 + H)} ~~~ q \in [\min(q_{\text{r}}, \zeta q_{\text{l}}), \zeta q_{\text{l}}], \\
  0 ~~~ \text{elsewhere},
  \end{cases}
\end{equation}
%\end{linenomath*}
where $C_0$ is a proportionality constant, $q_{\text{l}}$, $q_{\text{r}}$, and $\zeta q_{\text{l}}$ are the respective lower cut-off, roll-off, and upper cut-off wavenumbers, $H$ is the Hurst exponent. The boundary of the half-space is a random, self-affine, isotropic, bandwidth-limited rough surface for a given magnification $\zeta$, and the contact pressure ($p > 0$) at an arbitrary location within the contact area is a random variable. The PDF of the random contact pressure is denoted by $P_0(p, \zeta)$. Assuming that the variation of random contact pressure $p > 0$ with respect to the magnification is a Markov process and adopting the no re-entry assumption, the evolution of $P_0(p, \zeta)$ with magnification $\zeta$ can be approximated by the diffusion equation \cite{Persson01}:
\begin{equation}\label{E:diffusion_equ}
\frac{\partial}{\partial \zeta} P_0(p, \zeta) = \frac{1}{2} \frac{\text{d} \text{Var}(p)}{\text{d} \zeta} \frac{\partial^2}{\partial p^2} P_0(p, \zeta),
\end{equation}
where $\text{Var}(p) = \langle (p - \bar{p})^2 \rangle $ is the variance of the contact pressure. Readers who are unfamiliar with Persson's theory or who struggle to understand it should refer to a recent tutorial \cite{xu2024persson} for a complete derivation of Eq. \eqref{E:diffusion_equ} starting from two fundamental assumptions. Taking the boundary conditions, $P_0(p = 0, \zeta) = P_0(p \to \infty, \zeta) = 0$, the closed-form solution of Eq. \eqref{E:diffusion_equ} is a Gaussian distribution, from which its mirror image about $p = 0$ is subtracted, 
%\begin{linenomath*}
\begin{equation}\label{eq:P0_solution} 
P_0(p, \zeta) = \frac{1}{\sqrt{2 \pi \text{Var}(p)}} \left\{ \exp \left[- \frac{(p - \bar{p})^2}{2 \text{Var}(p)}\right] - \exp \left[- \frac{(p + \bar{p})^2}{2 \text{Var}(p)}\right] \right\}. 
\end{equation}
%\end{linenomath*}
The relative contact area is \cite{Xu24Fudge}
%\begin{linenomath*}
\begin{equation}\label{eq:Area}
A^*(\bar{p}, \zeta) = \int_0^{\infty} P_0(p, \zeta) \text{d}p = \text{erf} \left( \frac{\bar{p}}{\sqrt{2 \text{Var}(p)}}\right).
\end{equation}
%\end{linenomath*}
Wang and Müser \cite{Wang17} gave an approximate form of $\text{Var}(p)$,
%\begin{linenomath*}
\begin{equation}
\frac{\text{d}}{\text{d} \zeta} \text{Var}(p) = S(\bar{p}, \zeta) \frac{\text{d}}{\text{d} \zeta} V_{\text{p}},
\end{equation}
%\end{linenomath*}
where the fudge factor $S(\bar{p}, \zeta)$ is  a fourth-order polynomial of $A^*$ \cite{Wang17, Xu24Fudge},    
%\begin{linenomath*}
\begin{equation}\label{eq:fudge_factor_WM17}
S(\bar{p}, \zeta) = \gamma - \frac{2}{9} \left[ A^*(\bar{p}, \zeta) \right]^2 + \left(\frac{11}{9} - \gamma \right) \left[ A^*(\bar{p}, \zeta) \right]^4,
\end{equation}
%\end{linenomath*}
where $\gamma = 5/9$, and $V_{\text{p}}$ is the variance of contact pressure when the rough surface is completely flattened \cite{xu2024persson},
%\begin{linenomath*}
\begin{equation}\label{eq:V_equ}
V_{\text{p}} = \frac{\pi}{2} (E^*)^2 \int_{q_{\text{l}}}^{\zeta q_{\text{l}}} q^3 C(q) \text{d} q,
\end{equation}
where $E^* = E/(1 - \nu^2)$ is the plane strain modulus depending on Young's modulus $E$ and Poisson's ratio $\nu$. When $\bar{p}$ is sufficiently large, the fudge factor approaches unity and no correction is needed. The numerical method for solving Eq. \eqref{eq:Area} is explicitly given by Xu et al. \cite{Xu24Fudge}. Persson showed that the elastic strain energy stored in the flattened rough surface can be formulated as \cite{Persson08,Xu24Fudge} 
\begin{equation}\label{eq:Uel_fudge_factor}
U_{\text{el}}(\bar{p}, \zeta) = \frac{\pi A_{\text{n}} E^* }{2} \int_{q_{\text{l}}}^{\zeta q_{\text{l}}} q^2 C(q) A^*(\bar{p}, q/q_{\text{l}}) S(\bar{p}, q/q_{\text{l}}) \text{d} q,
\end{equation} 
where $A_{\text{n}}$ is the nominal contact area. 
%\end{linenomath*}
\section{PDF of interfacial gap}\label{sec:PDF_gap}
\subsection{Convection--diffusion equation}
Assuming that the variation of the interfacial gap with the magnification is a Markov process, the evolution of $P(g, \zeta)$ with $\zeta$ can be characterized incrementally by the Chapman--Kolmogorov equation in a backward manners,
%\begin{linenomath*}
\begin{equation}\label{eq:CK_PDF_gap}
P(g, \zeta) = \int_0^{\infty} P(g, \zeta|g', \zeta + \Delta \zeta) P(g', \zeta + \Delta \zeta) \text{d}g',
\end{equation} 
where $\Delta \zeta > 0$. The PDF $P(g, \zeta)$ can be represented by a piecewise function:
\begin{equation}\label{eq:PDF_gap}
P(g, \zeta) = \delta(g) \left[ 1 - \bar{A}^*(\bar{p}, \zeta) \right] + \theta(g) P_0(g, \zeta), 
\end{equation}
where $\bar{A}^*(\bar{p}, \zeta)$ is the relative non-contact area, 
\begin{equation}\label{eq:Probability_conservation}
\bar{A}^*(\bar{p}, \zeta) = \int_0^{\infty} P_0(g, \zeta) \text{d}g = 1 - A^*(\bar{p}, \zeta), 
\end{equation}
$\theta(g)$ is the Heaviside function ($\theta(g > 0) = 1$ and $\theta(g < 0) = 0$), and $P_0(g, \zeta)$ is the PDF of the positive interfacial gap within the non-contact area. The probability conservation is strictly held by Eq. \eqref{eq:PDF_gap}. An out-of-contact location ($g' > 0$) at magnification $\zeta + \Delta \zeta$  may remain out-of-contact ($g > 0$) or turn to contact status ($g = 0$) at the ``past"\footnote{We make an analogy between the time-dependent stochastic process (e.g., Brownian motion) and the magnification-dependent process for rough surface contact} (lower) magnification $\zeta$. Therefore, the transition probability $P(g, \zeta|g', \zeta + \Delta \zeta)$ can be represented using a piecewise function for $g' > 0$:
\begin{equation}\label{eq:Transition_PDF_gap}
P(g, \zeta | g', \zeta + \Delta \zeta) = \delta(g) \left[ 1 - \bar{A}_{\text{t}}^*(g', \zeta, \Delta \zeta) \right] + \theta(g) P_0(g, \zeta |g', \zeta + \Delta \zeta), 
\end{equation}
where
\begin{equation}
\bar{A}_{\text{t}}^*(g', \zeta, \Delta \zeta) = \int_{0^+}^{\infty} P_0(g, \zeta | g', \zeta + \Delta \zeta) \text{d} g.
\end{equation}
The transition probability $P_0(g, \zeta|g', \zeta + \Delta \zeta)$ deduces to a Dirac delta function $\delta(g - g')$ when $\Delta \zeta \to 0$. The following asymptotic behavior of $\bar{A}_{\text{t}}^*$ when $\Delta \zeta \to 0$ will be used in the later derivation: 
\begin{equation}
\lim \limits_{\Delta \zeta \to 0} \bar{A}_{\text{t}}^*(g' > 0, \zeta, \Delta \zeta) \to 1.
\end{equation}

Let us revisit the no re-entry assumption in Persson's theory. It can be stated mathematically as
\begin{equation}\label{eq:no_reentry}
P(p, \zeta + \Delta \zeta |p' = 0, \zeta) = \delta(p), 
\end{equation}
which implies that an out-of-contact spot at the present magnification $\zeta$ can only remain out-of-contact at any ``future" magnification $\zeta + \Delta \zeta$ \cite{Dapp14}. An equivalent statement of the no re-entry assumption is that,  given a contacting spot at magnification $\zeta + \Delta \zeta$, it remains in contact at any ``past" magnification $\zeta$, i.e., 
\begin{equation}\label{eq:no_reentry_P}
P(g, \zeta|g' = 0, \zeta + \Delta \zeta) = \delta(g), 
\end{equation}
which is equivalent to
\begin{equation}\label{eq:no_reentry_P0}
P_0(g, \zeta|g' = 0, \zeta + \Delta \zeta) = 0.
\end{equation}
Let $g' = 0$ in Eq. \eqref{eq:Transition_PDF_gap}, the equality is held with Eq. \eqref{eq:no_reentry_P0}. Therefore, Eq. \eqref{eq:Transition_PDF_gap} is valid for both $g' = 0$ and $g' > 0$. Bemporad and Paggi \cite{bemporad2015optimization,borri1999scaling} have already applied this assumption to accelerate a boundary element method (BEM) for rough surface contact. They achieved this by utilizing the contact domain with the coarser surface representation at a lower magnification as an initial guess.

Replacing all PDFs in Eq. \eqref{eq:CK_PDF_gap} with their piecewise representations given in Eqs. \eqref{eq:PDF_gap} and \eqref{eq:Transition_PDF_gap}, we can obtain that
\begin{equation}\label{eq:CK_P0_gap}
P_0(g, \zeta) = \int_0^{\infty} P_0(g, \zeta|g', \zeta + \Delta \zeta) P_0(g', \zeta + \Delta \zeta) \text{d}g', 
\end{equation}
i.e., the variation of the interfacial gap in the non-contact area with a decreasing magnification is a Markov process. Moreover, the above replacement can result in the following relation:
\begin{equation}\label{eq:CK_Abar}
\bar{A}^*(\bar{p}, \zeta) = \int_0^{\infty} \bar{A}_{\text{t}}^*(g', \zeta, \Delta \zeta)  P_0(g', \zeta + \Delta \zeta) \text{d}g',
\end{equation}
which leads to $\bar{A}^*(\bar{p}, \zeta + \Delta \zeta) \geq \bar{A}^*(\bar{p}, \zeta)$ indicating that the relative non-contact area monotonically grows with magnification till unit value is reached. This conclusion is the same as that found by Xu et al. \cite{xu2024persson} that the relative contact area monotonically reduces with magnification till vanishing.

Let $f(g) = P_0(g - \Delta g, \zeta|g, \zeta + \Delta \zeta) P_0(g, \zeta + \Delta \zeta)$. Rewriting Eq. \eqref{eq:CK_P0_gap} as $P_0(g, \zeta) = \int_{-g}^{\infty} f(g + \Delta g) \text{d}\Delta g$, a differential equation can be derived by expanding $f(g + \Delta g)$ in a Taylor series around $g$: 
\begin{equation}\label{eq:CK_differential_gap_1}
\bar{A}_{\text{t}}^*(g, \zeta, \Delta \zeta) P_0(g, \zeta + \Delta \zeta) - P_0(g, \zeta) = \sum_{n=1}^{\infty} \frac{-1}{n!} \int_{-\infty}^{g} (\Delta g)^n \frac{\partial^n}{\partial g^n} \left[ P_0(g - \Delta g, \zeta | g, \zeta + \Delta \zeta) P_0(g, \zeta + \Delta \zeta) \right] \text{d} \Delta g.
\end{equation}
By mathematical induction, the integral on the right-hand side of Eq. \eqref{eq:CK_differential_gap_1} can be expanded as
\begin{align}\label{eq:CK_differential_gap_2}
&\int_{-\infty}^{g} (\Delta g)^n \frac{\partial^n}{\partial g^n} \left[ P_0(g - \Delta g, \zeta | g, \zeta + \Delta \zeta) P_0(g, \zeta + \Delta \zeta) \right] \text{d} \Delta g = \notag \\
&\frac{\partial^n}{\partial g^n} \int_{-\infty}^g (\Delta g)^n P_0(g - \Delta g, \zeta | g, \zeta + \Delta \zeta) P_0(g, \zeta + \Delta \zeta) \text{d}\Delta g - n (\Delta g)^n \frac{\partial^{n-1}}{\partial g^{n-1}} P_0(0, \zeta|g, \zeta + \Delta \zeta) P_0(g, \zeta + \Delta \zeta).
\end{align}
If we divide both sides of Eq. \eqref{eq:CK_differential_gap_1} by $\Delta \zeta$ and let $\Delta \zeta \to 0$, then $\bar{A}_{\text{t}}^*(g > 0, \zeta, \Delta \zeta) = 1$ and $\lim \limits_{\Delta \zeta \to 0} P_0(0, \zeta|g > 0, \zeta + \Delta \zeta) = 0$. Substituting Eq. \eqref{eq:CK_differential_gap_2} into Eq. \eqref{eq:CK_differential_gap_1} divided by $\Delta \zeta \to 0$, we obtain the Kramers--Moyal expansion: 
\begin{equation}\label{E:Backward_KM_expansion}
\frac{\partial}{\partial \zeta} P_0(g, \zeta) = \sum_{n=1}^{\infty} \frac{(-1)^n}{n!} \frac{\partial^n}{\partial g^n} \left[B_n(g, \zeta) P_0(g, \zeta)\right],
\end{equation}
where  
\begin{equation}\label{E:Bn_forward_KM_expansion}
    B_n(g, \zeta) = \lim_{\Delta \zeta \to 0} \left( -\frac{1}{\Delta \zeta} \right) \int_{-\infty}^{g} (-\Delta g)^n P_0(g - \Delta g, \zeta | g, \zeta + \Delta \zeta) \text{d} \Delta g.
\end{equation}
The coefficient $B_n$ is the rate of change of the $n^{\text{th}}$ moment of the  interfacial gap with respect to $\zeta$. Adopting the approximation $B_n(g, \zeta) \approx B_n(\zeta) = \int_{0}^{\infty} B_n(g, \zeta) P_0(g, \zeta) \text{d}g$, and only keeping   terms up to the second order based on Pawula's lemma \cite{Pawula67, xu2024persson}, we obtain a convection--diffusion equation for $P_0(g, \zeta)$: 
\begin{equation}\label{eq:convec_diff_equation}
\frac{\partial}{\partial \zeta} P_0(g, \zeta) = -B_1(\zeta)  \frac{\partial}{\partial g} P_0(g, \zeta) + \frac{1}{2} B_2(\zeta) \frac{\partial^2}{\partial g^2} P_0(g, \zeta),
\end{equation}
where $B_1(\zeta)$ and $B_2(\zeta)$ are the drift and diffusion coefficients, respectively. 
%\end{linenomath*}
\subsection{Drift coefficient}
%\begin{linenomath*}
The drift coefficient represents the rate of change of the mean gap $\langle g \rangle$ with respect to $\zeta$ (see Appendix C in Ref. \cite{xu2024persson} for a complete derivation of a similar drift coefficient for contact pressure):
\begin{equation}\label{eq:B1}
B_1(\zeta) = \frac{\text{d} \langle g \rangle}{\text{d} \zeta}. 
\end{equation}
%\end{linenomath*}
The drift coefficient is strictly positive for a finite value of $\bar{p}$. Persson provided a thermal dynamic formulation for the mean gap \cite{Persson07}, 
%\begin{linenomath*}
\begin{equation}\label{eq:drift_coef_form1}
\langle g \rangle = \frac{1}{A_{\text{n}}} \int_{\bar{p}}^{\infty} \frac{\partial U_{\text{el}}}{\partial \bar{p}'} \frac{1}{\bar{p}'} \text{d} \bar{p}'.
\end{equation}
%\end{linenomath*}
Substituting Eqs. \eqref{eq:fudge_factor_WM17} and \eqref{eq:Uel_fudge_factor} into Eq. \eqref{eq:drift_coef_form1}, we have
%\begin{linenomath*}
\begin{equation}\label{eq:mean_gap}
\langle g \rangle = \frac{\pi E^*}{2} \int_{q_{\text{l}}}^{\zeta q_{\text{l}}} q^2 C(q) I(\bar{p}, q) \text{d} q, 
\end{equation}
%\end{linenomath*}
where 
%\begin{linenomath*}
\begin{equation}\label{eq:I_form}
I(\bar{p}, q) = \int_{\bar{p}}^{\infty} \frac{\partial}{\partial \bar{p}'} A^*(\bar{p}', \zeta) \left\{ \gamma - \frac{2}{3} \left[A^*(\bar{p}', \zeta) \right]^2 + 5 \left( \frac{11}{9} - \gamma \right) \left[A^*(\bar{p}', \zeta)\right]^4 \right\} \frac{1}{\bar{p}'} \text{d} \bar{p}'. 
\end{equation}
%\end{linenomath*}
To reduce the complexity of $I(\bar{p}, q)$, $A^*(\bar{p}, \zeta)$ in Eq. \eqref{eq:I_form} is simplified to an approximated form \cite{Almqvist11,Afferrante18}, 
%\begin{linenomath*}
\begin{equation}
A^*(\bar{p}, \zeta) = \text{erf} \left( \frac{\bar{p}}{\sqrt{2 V_{\text{p}}(\zeta)}} \right), 
\end{equation}
and 
\begin{equation}
\frac{\partial}{\partial \bar{p}} A^*(\bar{p}, \zeta) = \sqrt{\frac{2}{\pi V_{\text{p}}(\zeta)}} \exp\left( -\frac{\bar{p}^2}{2 V_{\text{p}}(\zeta)} \right). 
\end{equation}
%\end{linenomath*}
Substituting Eq. \eqref{eq:mean_gap} into Eq. \eqref{eq:B1}, an explicit form of $B_1$ can be obtained. In Section 4, the numerical method for solving Eq. \eqref{eq:convec_diff_equation} requires the discretization of Eq. \eqref{eq:B1} using the Crank--Nicolson implicit scheme. The drift coefficient in Eq. \eqref{eq:B1} is approximated by the backward difference scheme. Therefore, it is not necessary to derive the explicit form of $B_1$. 
\subsection{Diffusion coefficient}
The diffusion coefficient represents the rate of change of the gap variance $\langle g^2 \rangle$ with respect to $\zeta$ (see Appendix C in Ref. \cite{xu2024persson} for a complete derivation of a similar diffusion coefficient for contact pressure):
%\begin{linenomath*}
\begin{equation}\label{eq:B2}
B_2(\zeta) = \frac{\text{d} \langle g^2 \rangle}{\text{d} \zeta}. 
\end{equation}
%\end{linenomath*}
The diffusion coefficient is strictly positive for a finite value of $\bar{p}$. Since $\langle g^2 \rangle = \langle g \rangle^2 + \text{Var}(g)$, we can rewrite Eq. \eqref{eq:B2} as
%\begin{linenomath*}
\begin{equation}\label{eq:B2_1}
B_2(\zeta) = 2 \langle g \rangle \frac{\text{d} \langle g \rangle}{\text{d} \zeta} + \frac{\text{d} \text{Var}(g)}{\text{d} \zeta}. 
\end{equation}
%\end{linenomath*}
When $A^*$ is vanishing, $\text{Var}(g) \approx V_{\text{h}}(\zeta)$, where $V_{\text{h}}(\zeta)$ is the variance of the rough surface height at magnification $\zeta$. According to Eq. \eqref{eq:mean_gap}, $\text{d} \langle g \rangle/\text{d} \zeta \to  0$ once $A^* \to 0$. Thus, $B_2$ may be approximated by its asymptotic behavior when $A^* \to 0$,  
%\begin{linenomath*}
\begin{equation}\label{eq:B2_1}
B_2(\zeta) = \frac{\text{d} V_{\text{h}}}{\text{d} \zeta}, 
\end{equation}
%\end{linenomath*}
where $V_{\text{h}}$ has an integral form in terms of $C(q)$\cite{xu2024persson}: 
%\begin{linenomath*}
\begin{equation}
V_{\text{h}} = 2 \pi \int_{q_{\text{l}}}^{\zeta q_{\text{l}}} q C(q) \text{d}q.
\end{equation}
%\end{linenomath*}
\section{Finite difference method}\label{sec:FDM}
The convection--diffusion equation is a typical initial-boundary value problem. When $\zeta = 1$, the rough surface reduces to a perfectly flat surface, and it has a conformal contact with the rigid flat over the entire interface. Thus, the initial condition of Eq. \eqref{eq:convec_diff_equation} is
%\begin{linenomath*}
\begin{equation}\label{eq:Initial_condition}
P_0(g > 0, \zeta = 1) = 0. 
\end{equation}
%\end{linenomath*}
Besides the vanishing boundary condition ($P_0(g \to \infty, \zeta) = 0$), by applying $\displaystyle{\int_{0}^{\infty} g \times\cdots \text{d}g}$ to both sides of Eq. \eqref{eq:convec_diff_equation}, we can obtain the boundary condition at $g = 0$ as
%\begin{linenomath*}
\begin{equation}\label{eq:Boundary_condition}
P_0(g = 0, \zeta) = 2 A^*(\bar{p}, \zeta) B_1(\zeta)/B_2(\zeta).
\end{equation}
%\end{linenomath*}
 In general, the convection--diffusion equation (Eq. \eqref{eq:convec_diff_equation}) with   initial and boundary conditions (Eqs. \eqref{eq:Initial_condition} and \eqref{eq:Boundary_condition}) cannot be solved analytically. The finite difference method is commonly adopted to solve the convection--diffusion equation in   statistical mechanics \cite{Chang70, Park96}. However, those solvers require probability conservation (i.e., $\int_0^{\infty} P_0(g, \zeta) \text{d}g = 1$), which is not satisfied in the present work. Hence, we develop a new finite difference method for solving  the non-conserved $P_0(g, \zeta)$.

Let the computational domain be $\Omega = \{(\zeta', g)| \zeta' \in [1, \zeta], g \in [0, g_{\max}] \}$. The gap axis is uniformly discretized with a constant mesh interval of $\Delta_{\text{g}} = g_{\max}/(n_{\text{g}} - 1)$, so that $g_i = (i-1) \Delta_{\text{g}}$, $i = 1, \cdots, n_{\text{g}}$. The magnification axis is discretized non-uniformly into a monotonic sequence ${\boldsymbol \zeta} = [\zeta_0, \zeta_1, \cdots, \zeta_{n_{\zeta}}]$ with $\zeta_0 = 1$, $\zeta_{n_{\zeta}} = \zeta$, and $(\zeta'_{i+1} - \zeta'_i)/(\zeta'_i - \zeta'_{i-1}) = C_1$, where $C_1$ is a constant. The PDF value at grid point $(g_i, \zeta'_j)$ is denoted by $P_i^j$.

\subsection{Implicit scheme}
The convection-diffusion equation is discretized using the implicit Crank--Nicolson scheme \cite{press2007numerical}: 
%\begin{linenomath*}
\begin{align}\label{eq:FK_discrete_1}
\frac{P_i^j - P_i^{j-1}}{\zeta'_j - \zeta'_{j-1}} =  - &\frac{1}{2} \frac{\langle g \rangle(\zeta_j') - \langle g \rangle(\zeta_{j-1}')}{\zeta_j' - \zeta_{j-1}'} \left[ \frac{P_{i + 1}^j - P_{i-1}^j}{2 \Delta_{\text{g}}} + \frac{P_{i + 1}^{j-1} - P_{i-1}^{j-1}}{2 \Delta_{\text{g}}} \right] + \notag \\
&\frac{1}{4} \frac{V_{\text{h}}(\zeta_j') - V_{\text{h}}(\zeta_{j-1}')}{\zeta_j' - \zeta_{j-1}'} \left[ \frac{P_{i+1}^j - 2 P_i^j + P_{i-1}^j}{(\Delta_{\text{g}})^2} + \frac{P_{i+1}^{j-1} - 2 P_i^{j-1} + P_{i-1}^{j-1}}{(\Delta_{\text{g}})^2} \right],
\end{align}
%\end{linenomath*}
where partial derivatives with respect to $\zeta$ and $g$ are approximated using the backward and central difference, respectively. This implicit scheme results in   unconditional stability. Reorganizing Eq. \eqref{eq:FK_discrete_1}, we can get a tridiagonal system of equations:
%\begin{linenomath*}
\begin{equation}\label{eq:FK_discrete_2}
-A P_{i-1}^j + B P_i^j - C P_{i+1}^j = R_i,
\end{equation}
where
\begin{align}
\Delta_{\zeta} &= \zeta'_j - \zeta'_{j-1}, ~~~ a_j = \frac{\langle g \rangle(\zeta_j') - \langle g \rangle(\zeta_{j-1}')}{\Delta_{\zeta}},
~~~ v_j = \frac{1}{2} \frac{V_{\text{h}}(\zeta_j') - V_{\text{h}}(\zeta_{j-1}')}{\Delta_{\zeta}}, \notag \\
A &= \frac{1}{4 \Delta_{\text{g}}} \left( a_j +  \frac{2 v_j}{\Delta_{\text{g}}} \right), ~~~ B = \frac{1}{\Delta_{\zeta}} + \frac{v_j}{(\Delta_{\text{g}})^2}, ~~~  C = \frac{1}{4 \Delta_{\text{g}}} \left( - a_j + \frac{2 v_j}{\Delta_{\text{g}}} \right),\notag \\
R_i &= A P_{i-1}^{j-1} + \left( \frac{1}{\Delta_{\zeta}} - A - C \right) P_i^{j-1} + C P_{i+1}^{j-1}. \label{eq:Coefs}
\end{align}
%\end{linenomath*}
\subsection{Initial and boundary conditions}
The discretized form of the initial condition in Eq. \eqref{eq:Initial_condition} is
%\begin{linenomath*}
\begin{equation}
P_{i}^{j = 0} = 0 ~~~ \forall i.
\end{equation}
%\end{linenomath*}
If $g_{\max}$ is sufficiently large, the vanishing boundary condition $P_0(g \to \infty, \zeta) = 0$ can be used over all boundary nodes: 
%\begin{linenomath*}
\begin{equation}
P_{i = n_{\text{g}}}^j = 0 ~~~ \forall j.
\end{equation}
%\end{linenomath*}
However, Eq. \eqref{eq:Boundary_condition} cannot strictly guarantee probability conservation over the discretized domain. Let us first discretize Eq. \eqref{eq:Probability_conservation} over the discretized domain:  
%\begin{linenomath*}
\begin{equation}\label{eq:BC2_discrete_1}
A^*(\bar{p}, \zeta'_j) + \sum_{i = 1}^{n_{\text{g}}} P_i^j \Delta_{\text{g}} = 1. 
\end{equation}
%\end{linenomath*}
Summing Eq. \eqref{eq:FK_discrete_2} with $i = 2, \cdots, n_{\text{g}}-1$, utilizing Eq. \eqref{eq:BC2_discrete_1} and the approximation $P_{n_{\text{g}}-1}^j = 0$ $\forall j$:
%\begin{linenomath*}
\begin{equation}\label{eq:BC2_discrete}
\left(B - C \right) P_1^j - C P_2^j = \frac{1}{\Delta_{\zeta} \Delta_{\text{g}}} \left[ A^*(\bar{p}, \zeta'_{j-1})) - A^*(\bar{p}, \zeta'_j) \right] + \left( \frac{1}{\Delta_{\zeta}} - A \right) P_1^{j-1} + C P_2^{j-1},
\end{equation}
%\end{linenomath*}
which strictly guarantees probability conservation. 
Assembling Eq. \eqref{eq:BC2_discrete} and Eq. \eqref{eq:FK_discrete_2} with $i = 2, \cdots, n_{\text{g}} - 1$, an alternative triagonal system of equations is obtained: 
%\begin{linenomath*}
\begin{align}\label{eq:triag_system_1}
&
\begin{bmatrix}
B - C & -C & 0          &  \dots               & 0 & 0              & 0      \\
-A & B & -C &                   & 0   & 0              & 0      \\
0 &  -A & B &                & 0   & 0              & 0        \\
\vdots     &            &            & \ddots     &          &                & \vdots      \\
0          & 0          &      0 &      &        -A & B & -C \\
0           & 0         & 0 &  \dots          &  0              & -A   & B
\end{bmatrix}
\begin{Bmatrix}
P_1^j \\
P_2^j \\
P_3^j \\
\vdots
\\
P_{n_{\text{g}}-2}^j \\
P_{n_{\text{g}}-1}^j
\end{Bmatrix}
= \notag \\
&
\begin{Bmatrix}
\displaystyle{\frac{1}{\Delta_{\zeta} \Delta_{\text{g}}} \left[ A^*(\bar{p}, \zeta_{j-1})) - A^*(\bar{p}, \zeta_j) \right] + \left( \frac{1}{\Delta_{\zeta}} - A \right) P_1^{j-1} + C P_2^{j-1}} \\
R_2 \\
R_3 \\
\vdots
\\
R_{n_{\text{g}}-2} \\
R_{n_{\text{g}}-1}
\end{Bmatrix}.
\end{align}
%\end{linenomath*}
\subsection{Nonnegativity criterion}
To find the criterion under which $P_0(g, \zeta) \geq 0$ is strictly satisfied, let us first simplify Eq. \eqref{eq:triag_system_1} as follows:
%\begin{linenomath*}
\begin{equation}\label{eq:triag_system_2}
-A_i P_{i-1}^j + B_i P_i^j - C_i P_{i+1}^j = \bar{R}_i, ~~~ i = 1, \cdots, n_{\text{g}} - 1,
\end{equation}
%\end{linenomath*}
where
%\begin{linenomath*}
\begin{align}
A_i &=
\begin{cases}
0 ~~~~ i = 1\\
A ~~~ i > 1
\end{cases}
, ~~~ B_i = 
\begin{cases}
B - C ~~~ i = 1\\
B ~~~ i > 1
\end{cases},
~~~ C_i =
\begin{cases}
C ~~~ i < n_{\text{g}} - 1 \\
0 ~~~~ i = n_{\text{g}} - 1
\end{cases},
\notag \\
\bar{R}_i &=
\begin{cases}
\displaystyle{\frac{1}{\Delta_{\zeta} \Delta_{\text{g}}} \left[ A^*(\bar{p}, \zeta'_{j-1})) - A^*(\bar{p}, \zeta'_j) \right] + \left( \frac{1}{\Delta_{\zeta}} - A \right) P_1^{j-1} + C P_2^{j-1}} ~~~ i = 1 \\
R_i ~~~ i > 1
\end{cases}
.
\end{align}
%\end{linenomath*}
Chang and Cooper provided a sufficient but not necessary condition \cite{Chang70, Park96}:
%\begin{linenomath*}
\begin{subequations}
\begin{align}
B_i \geq A_i + C_i, \label{eq:Non_negativity_1}\\
A_i, ~B_i, ~C_i \geq 0, \label{eq:Non_negativity_2}\\
\bar{R}_i \geq 0, \label{eq:Non_negativity_3}
\end{align}
\end{subequations}
%\end{linenomath*}
which guarantees $P_0(g, \zeta) \geq 0$. Based on the identity $B - A - C = 1/\Delta_{\zeta}$, the first condition in Eq. \eqref{eq:Non_negativity_1}   holds $\forall i$. Since $A$ and $B$ are strictly positive with a finite value of $\bar{p}$, the second condition in Eq. \eqref{eq:Non_negativity_2} requires $C \geq 0$. For $i > 1$, $C \leq 1/\Delta_{\zeta} - A$ can guarantee $\bar{R}_i = R_i \geq 0$. Using the tight constraint $0 \leq C \leq 1/\Delta_{\zeta} - A$ and the fact that $A^*(\bar{p}, \zeta_{j-1}') > A^*(\bar{p}, \zeta_{j}')$, we have $\bar{R}_1 = R_1 \geq 0$. Consequently, Eqs. \eqref{eq:Non_negativity_1}, \eqref{eq:Non_negativity_2}, and \eqref{eq:Non_negativity_3} are satisfied if
%\begin{linenomath*}
\begin{equation}\label{eq:delta_g_criterion}
\Delta_{\text{g}} \in \left[\sqrt{v_j \Delta_{\zeta}}, ~2v_j/a_j  \right].
\end{equation}
%\end{linenomath*}
Eq. \eqref{eq:delta_g_criterion} may result in different $\Delta_{\text{g}}$ at different magnifications. To solve this inconvenience, we provide a looser constraint for $\Delta_{\text{g}}$:
%\begin{linenomath*} 
\begin{equation}\label{eq:delta_g_criterion_1}
\Delta_{\text{g}} = \frac{1}{2}\max_{\forall j} \left( \frac{2 v_j}{a_j} \right).
\end{equation}
%\end{linenomath*}
Based on our numerical experience, the above criterion can strictly satisfy the nonnegativity criterion. 

\section{Results and discussion}\label{sec:RandD}

We investigate the interaction between the rigid flat and a specific group of rough surfaces, where $q_{\text{l}} = q_{\text{r}} = 10$ (1/mm), $\zeta = 100$, $H = 0.8$, $\sqrt{V_{\text{h}}(\zeta)} = 6 \times 10^{-3}$ (mm). According to Eq. (A.10) in Appendix A of Ref. \cite{xu2024persson}, the constant $C_0$ in Eq. \eqref{E:PSD} can be determined by
%\begin{linenomath*}
\[
C_0 = \frac{V_{\text{h}} H}{\pi \left(q_{\text{l}}^{-2H} - \zeta^{-2H} q_{\text{l}}^{-2H}\right)}.
\]
%\end{linenomath*}
Reference solutions are provided by GFMD \cite{Zhou20, Zhou19}. Other numerical methods, e.g., boundary element method \cite{bemporad2015optimization, frerot2020tamaas}, can provide almost identical results. Self-affine rough surface topography used in GFMD is generated artificially. Two major approaches are commonly adopted to generate the self-affine random rough surfaces: random midpoint displacement (RMD) \cite{meakin1998fractals,paggi2011contact} and spectral synthesis \cite{hu1992simulation,Xu17}. The latter one is used in the current study to guarantee the generated random rough surface topography is strictly band-width limited, isotropic, periodic, and smooth. One rough surface with the prescribed PSD given in Eq. \eqref{E:PSD} is numerically generated with $8,192 \times 8,192$ sampling points over a nominal contact area $A_{\text{n}} = \pi^2$ (mm$^2$). Besides the present model, three other models are compared: \texttt{Almqvist_etal_11}\cite{Almqvist11}, \texttt{Afferrante_etal_18}\cite{Afferrante18}, \texttt{Joe_etal_23}\cite{Joe23}. Detailed expressions for the first two models can be found in Appendix A. 

\begin{figure}[h!]
  \centering
  \includegraphics[width=9cm]{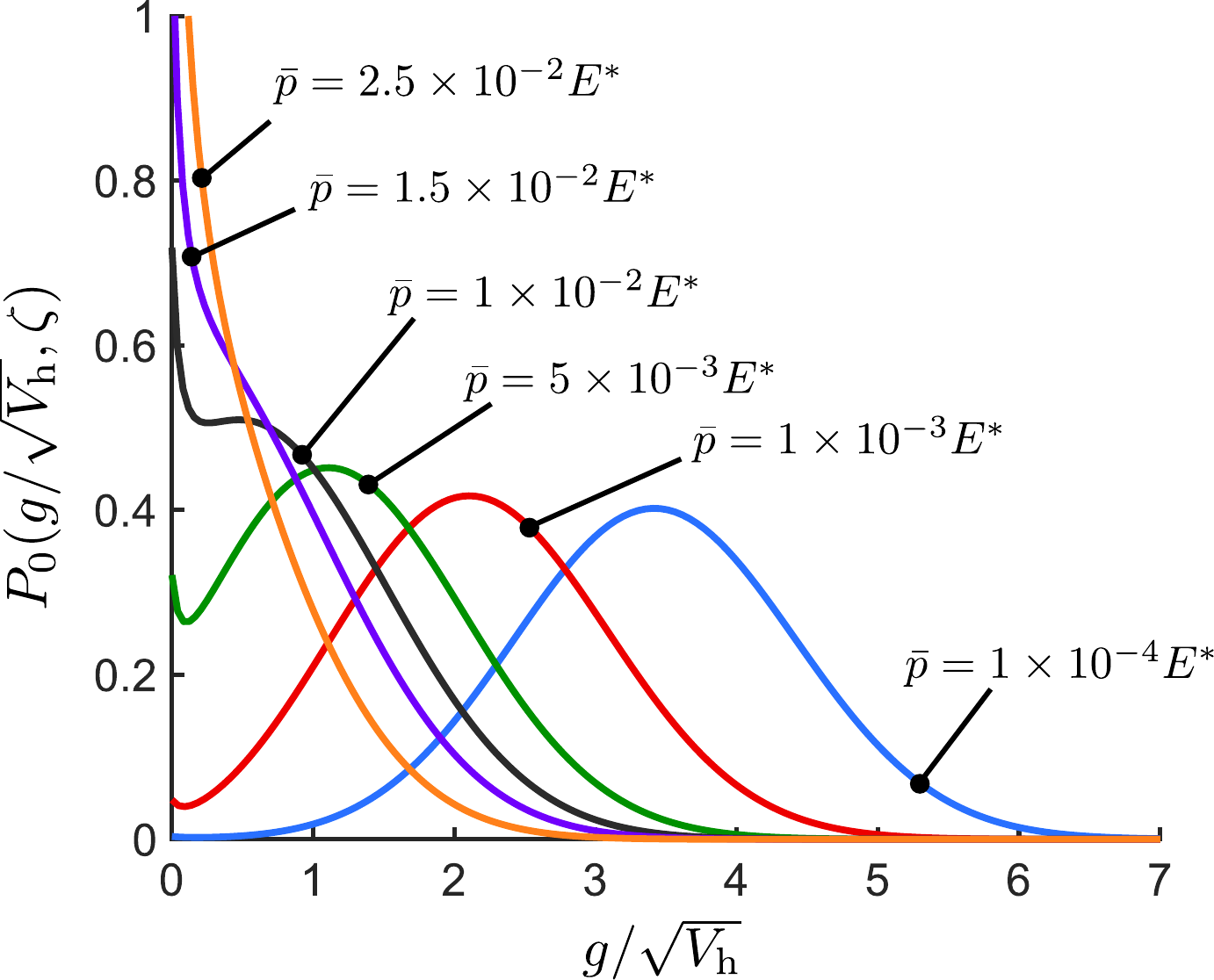}
  %\internallinenumbers 
  \caption{Evolution of $P_0(g/\sqrt{V_{\text{h}}}, \zeta)$ with respect to $\bar{p}/E^*$, solved by   present work. Values of   relative contact area $A^*$ with six values of $\bar{p}/E^*$ (in ascending order) are $7.664 \times 10^{-4}$, $7.664 \times 10^{-3}$, $3.834 \times 10^{-2}$, $7.678 \times 10^{-2}$, $0.115$, and $0.191$.}\label{fig:Fig_2}
\end{figure}

\begin{figure}[h!]
  \centering
  %\internallinenumbers 
  \includegraphics[width=14cm]{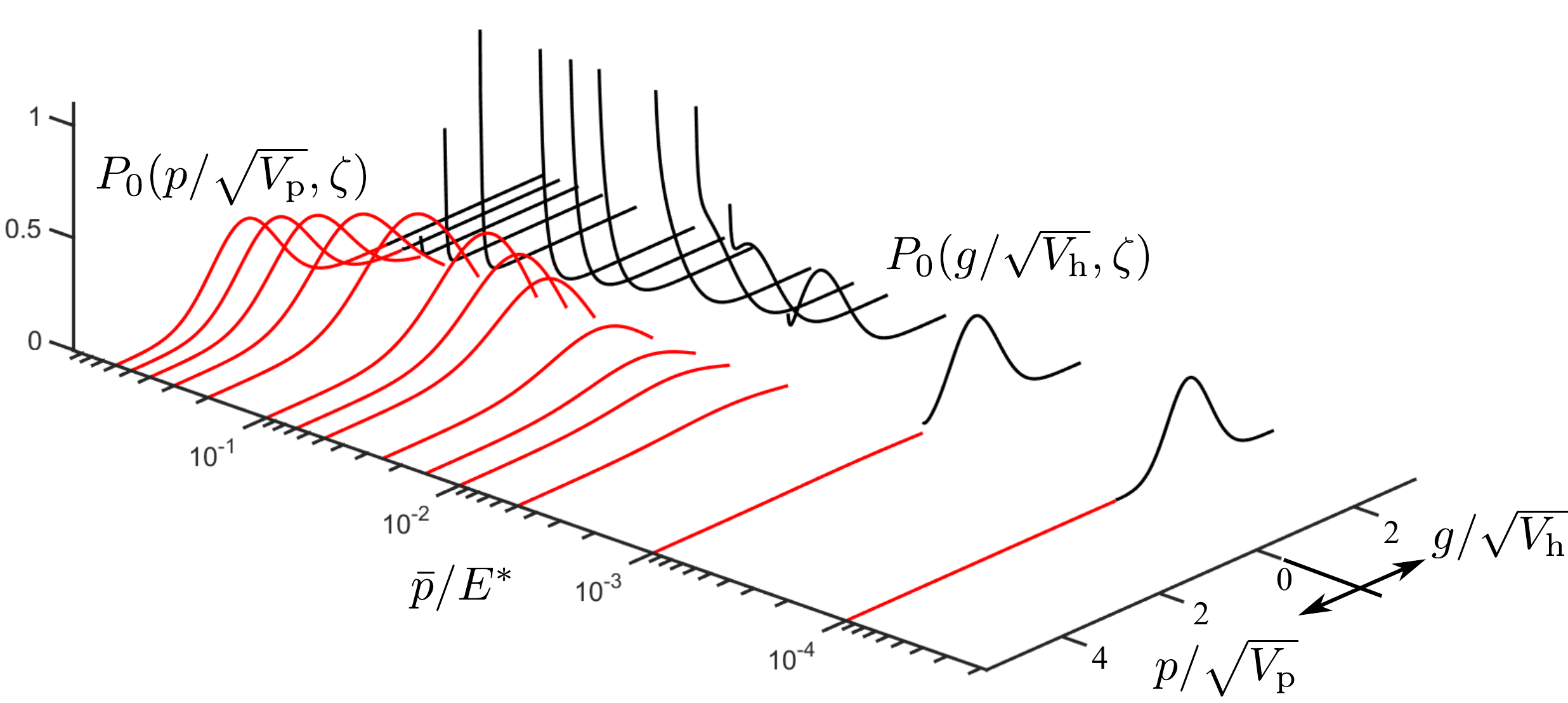}
  \caption{Co-evolution of $P_0(p/\sqrt{V_{\text{p}}}, \zeta)$ and $P_0(g/\sqrt{V_{\text{h}}}, \zeta)$ with respect to $\bar{p}/E^*$. Both PDFs are solved by   present work.}\label{fig:Fig_3}
\end{figure}

When $\bar{p}$ is extremely low, the entire interface is nearly out-of-contact. In this extreme case, the drift coefficient $B_1$ vanishes, and the expression of the diffusion coefficient in Eq. \eqref{eq:B2_1}   strictly holds. Therefore, Eq. \eqref{eq:convec_diff_equation} has the following asymptotic Gaussian solution when $\bar{p} \to 0$: 
%\begin{linenomath*}
\begin{equation}
P_0(g, \zeta) \approx \frac{1}{\sqrt{2 \pi V_{\text{h}}(\zeta)}} \exp \left[ -\frac{(g - \langle g \rangle)^2}{2 V_{\text{h}}(\zeta)} \right]. 
\end{equation}
%\end{linenomath*}
As $\bar{p}$ monotonically increases from zero, $P_0(g, \zeta)$, with an initial shape of a Gaussian distribution, is gradually ``squeezed" to the lower gap range, as shown in Fig. \ref{fig:Fig_2}. Unlike the absorbing boundary condition of $P_0(p, \zeta)$, $P_0(g \to 0^+, \zeta)$ is generally positive. At medium load ranges (such as $\bar{p} = 5 \times 10^{-3} E^*$), we can observe a double-peaked PDF. As $\bar{p}$ further increases, two peaks in the PDF curve coalesce into a semi-impulse function. Fig. \ref{fig:Fig_3} shows the co-evolution of $P_0(g, \zeta)$ and $P_0(p, \zeta)$. We can clearly see how they ``leak" as $\bar{p}/E^*$ increases and decreases by more than three orders of magnitude.

\begin{figure}[h!]
  \centering
  %\internallinenumbers 
  \includegraphics[width=16cm]{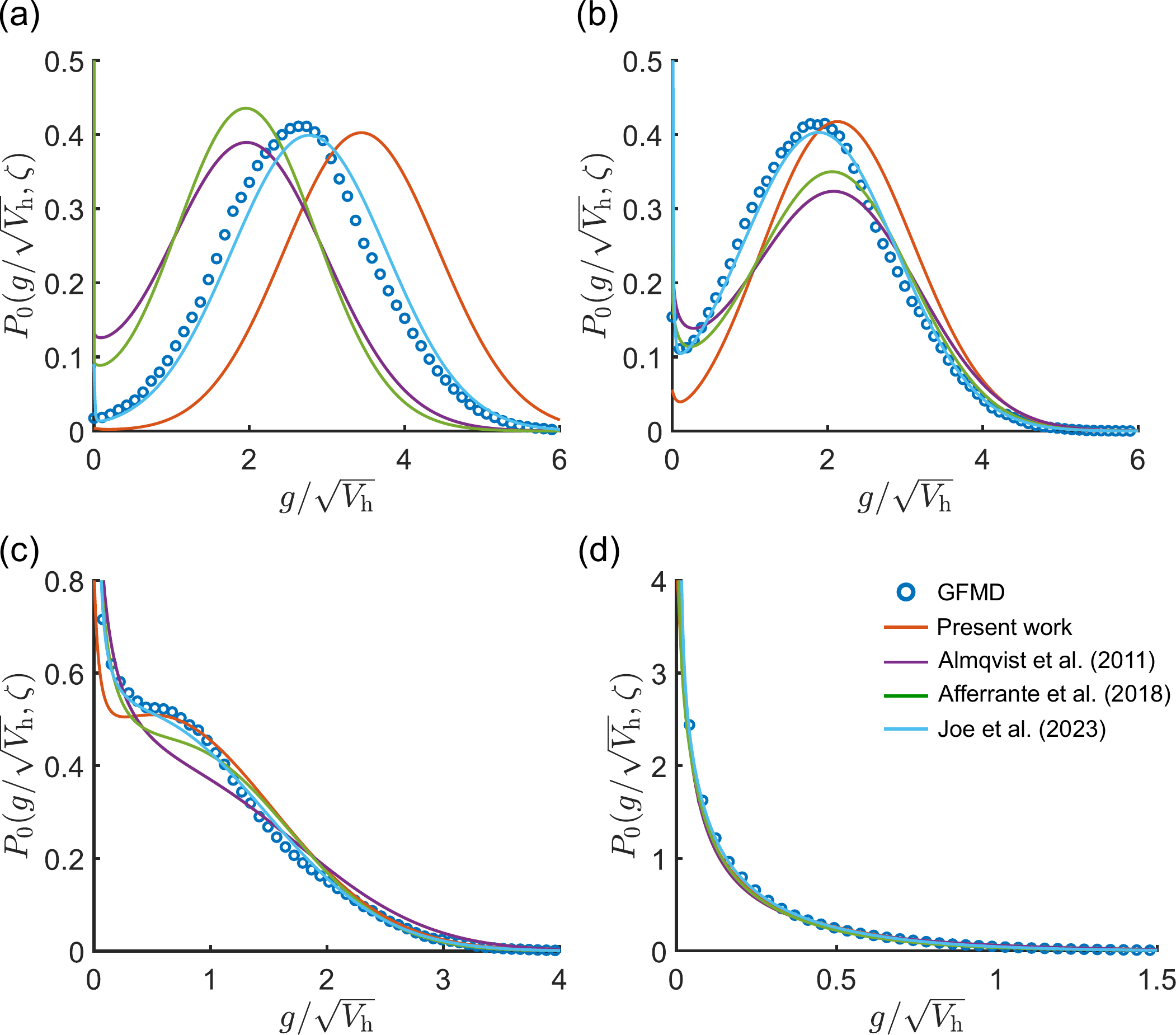}
  \caption{$P_0(g/\sqrt{V_{\text{h}}}, \zeta)$ predicted by various models. (a) $\bar{p}/E^* = 1 \times 10^{-4}$; (b) $\bar{p}/E^* = 1 \times 10^{-3}$; (c) $\bar{p}/E^* = 1 \times 10^{-2}$; (d) $\bar{p}/E^* = 6.5 \times 10^{-2}$.}\label{fig:Fig_4}
\end{figure}

Fig. \ref{fig:Fig_4} compares the predicted $P_0(g, \zeta)$ under four   values of $\bar{p}$. When $\bar{p}/E^* = 1 \times 10^{-4}$, the non-contact area occupies nearly the entire interface ($A^* = 7.664 \times 10^{-4}$). \texttt{Joe_etal_23} has the best agreement with GFMD. The PDF predicted by the present work has a slight offset from those predicted by GFMD and \texttt{Joe_etal_23} toward the larger gap ranges. The PDFs predicted by \texttt{Almqvist_etal_11} and \texttt{Afferrante_etal_18} are slightly shifted from GFMD results toward the lower gap ranges. As $\bar{p}/E^*$ increases to $1 \times 10^{-3}$, the non-contact area is still dominant over the interface ($A^* = 7.664 \times 10^{-3}$). A double-peaked PDF is captured by all models. \texttt{Joe_etal_23} still has the best agreement with GFMD. At the lower gap ranges, the improved Persson's theory (\texttt{Afferrante_etal_18}) and  \texttt{Joe_etal_23} accurately recreate the peak at $g = 0$, while the original Persson's theory (\texttt{Almqvist_etal_11}) and the present work, respectively, overestimate and underestimate this peak. At medium gap ranges, the present work shows  better accuracy than \texttt{Almqvist_etal_11} and \texttt{Afferrante_etal_18}. At the larger gap ranges, nearly all theoretical models are almost identical with GFMD, while the PDF is slightly overestimated by the present work. As $\bar{p}/E^*$ further increases to $1 \times 10^{-2}$, the contact area starts to become visible ($A^* = 0.0744$). \texttt{Joe_etal_23} retains a nearly perfect agreement with GFMD. The present work shows good agreement with GFMD at the higher pressure ranges. \texttt{Afferrante_etal_18} has a nearly identical prediction to that by GFMD, while the present work and \texttt{Almqvist_etal_11}, respectively, underestimate and slightly overestimate the PDF at the lower pressure ranges. When $\bar{p}/E^*$ eventually increases to $6.5 \times 10^{-2}$, the contact area occupies nearly half of the interface ($A^* = 0.4193$). The PDF data predicted by all models are nearly identical over the investigated gap ranges. 

\begin{figure}[h!]
  \centering
  \includegraphics[width=16cm]{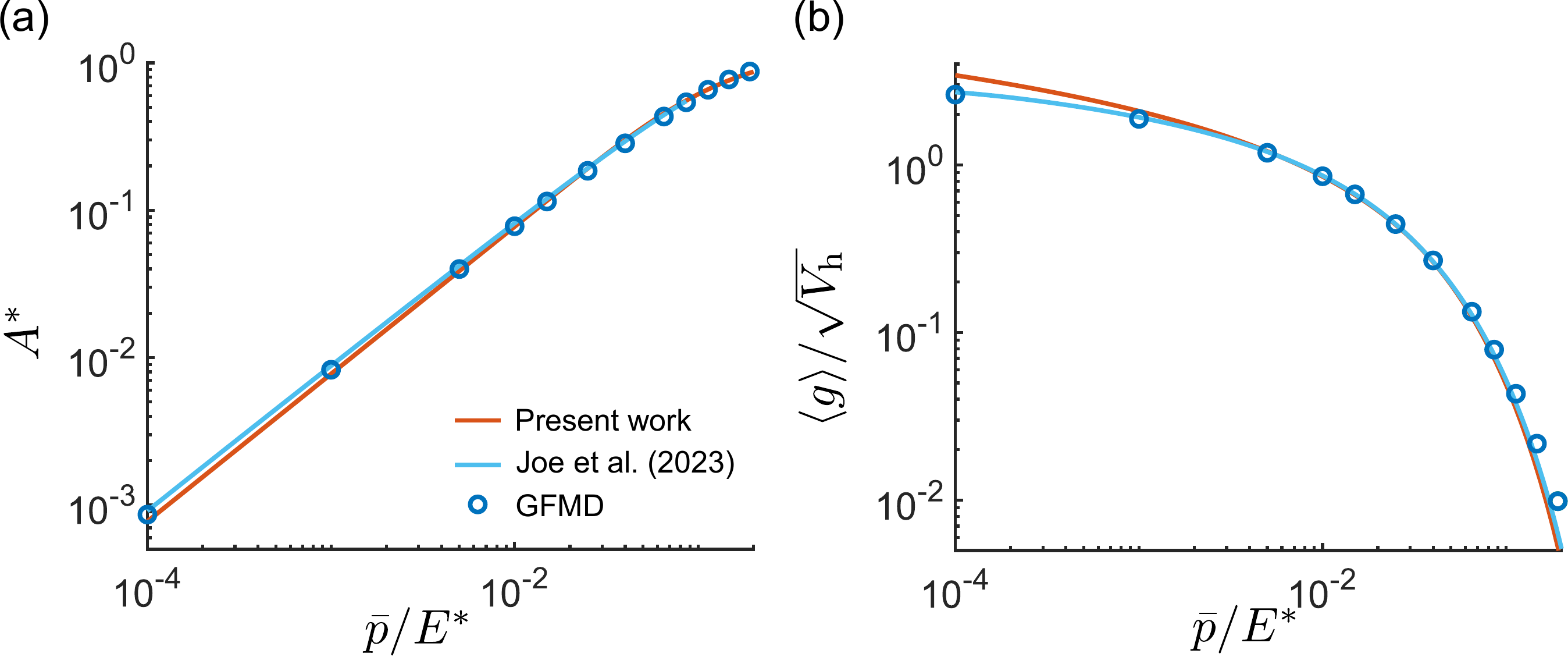}
  %\internallinenumbers 
  \caption{Predictions by present work, \texttt{Joe_etal_23}, and GFMD. (a) Variation of relative contact area with $\bar{p}/E^*$; (b) Variation of dimensionless average interfacial gap $\langle g \rangle/\sqrt{V_{\text{h}}}$ with $\bar{p}/E^*$.}\label{fig:Fig_5} 
\end{figure}

Assuming the convection-diffusion equation, given in Eq. \eqref{eq:convec_diff_equation}, accurately describes the evolution of $P_0(g, \zeta)$ with respect to magnification $\zeta$, the accuracy of $P_0(g, \zeta)$ predicted by the present work is governed by the accuracy of $A^*(\bar{p}, \zeta)$, the diffusion and drift coefficients. When $\bar{p}$ is extremely low (such as Fig. \ref{fig:Fig_4}(a,b)), the diffusion coefficient given by Eq. \eqref{eq:B2_1} is strictly satisfied. Since the drift coefficient in Eq. \eqref{eq:FK_discrete_1} is approximated by the backward difference of $\langle g \rangle(\zeta)$, the deviation shown in Fig. \ref{fig:Fig_4}(a,b) may be coupled with the accuracy of $A^*$ and $\langle g \rangle$. Fig. \ref{fig:Fig_5}(a) shows $A^*(\bar{p}/E^*, \zeta)$ relations predicted by the present work, \texttt{Joe_etal_23}, and GFMD, and all results are nearly identical throughout the entire investigated $\bar{p}$ ranges. This implies that the areas underneath all PDF curves shown in Fig. \ref{fig:Fig_4}(a--d) are nearly identical. Fig. \ref{fig:Fig_5}(b) illustrates $\langle g \rangle$ vs. $\bar{p}$ relations at magnification $\zeta$ predicted by the present work, \texttt{Joe_etal_23}, and GFMD. The present work overestimates $\langle g \rangle$ at low pressure ranges, while \texttt{Joe_etal_23} has  perfect agreement with GFMD. The overestimation of $\langle g \rangle$ by the present work is a key to explain its overall shift from GFMD and \texttt{Joe_etal_23} toward larger gap ranges, as shown in Fig. \ref{fig:Fig_4}(a-b). If we keep the area underneath $P_0(g, \zeta)$ the same and shift it as a whole to the lower gap ranges to match $\langle g \rangle$ as predicted by GFMD and \texttt{Joe_etal_23}, we can expect improved agreement. In the medium load ranges, the predicted $\langle g \rangle$ vs. $\bar{p}$ relations are all identical. Thus, the minor mismatch between the present work and GFMD in Fig. \ref{fig:Fig_4}(c) may be caused by an overestimated diffusion coefficient, which broadens the span of $P_0(g, \zeta)$ over the gap axis. In the high load ranges, $\langle g \rangle$ vs. $\bar{p}$, as  predicted by \texttt{Joe_etal_23} and the present work, are almost the same, but slightly lower than that by GFMD. This mean gap underestimation explains why the $P_0(g, \zeta)$ curves predicted by all theoretical models are slightly shifted to the lower gap ranges, see Fig. \ref{fig:Fig_4}(d). 

\begin{figure}[h!]
  \centering
  \includegraphics[width=16cm]{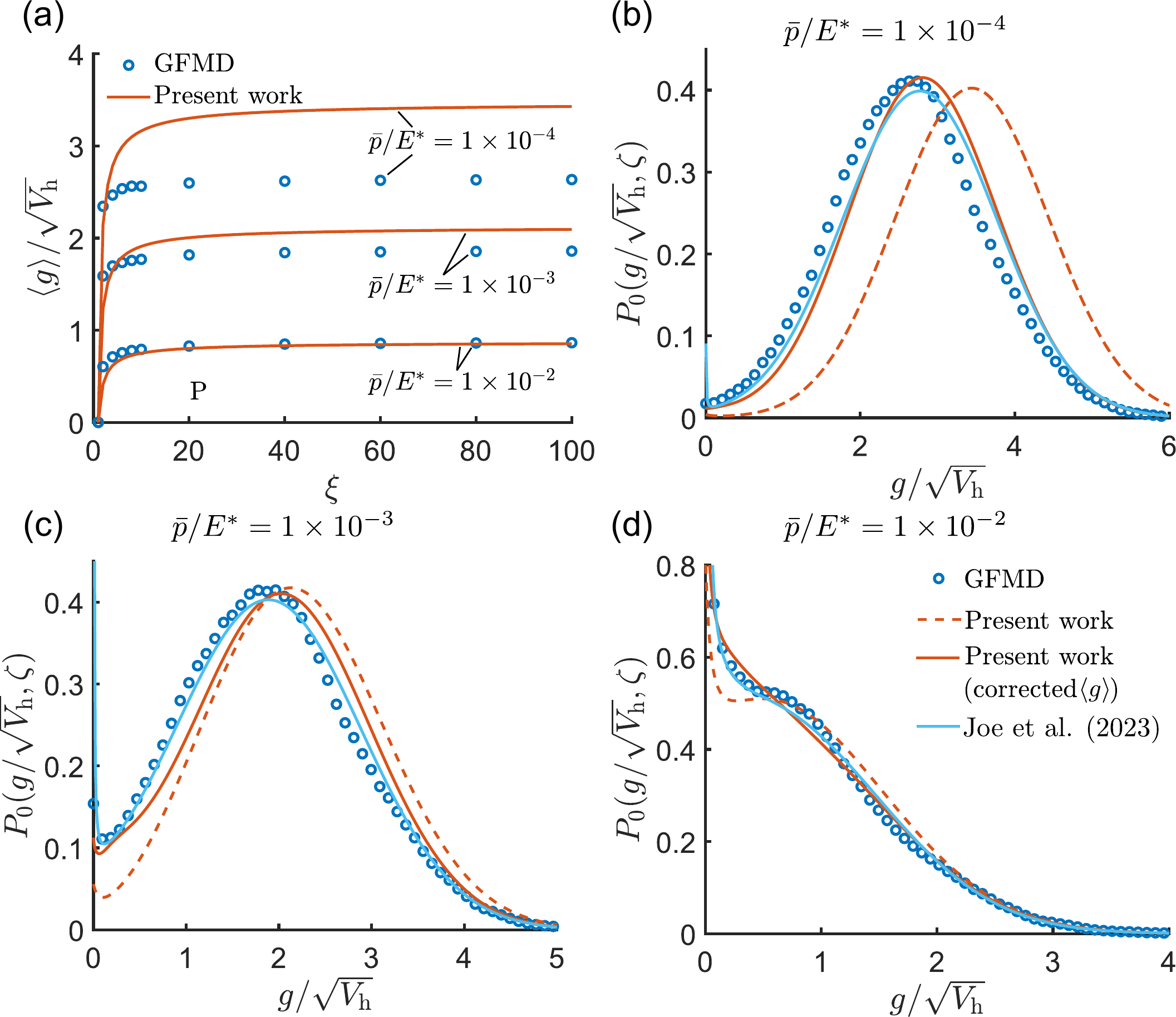}
  %\internallinenumbers 
  \caption{(a) Magnification-dependent mean gap, $\langle g \rangle$, under $\bar{p}/E^* = 1 \times 10^{-4}, 1 \times 10^{-3}$ and $1 \times 10^{-2}$, predicted by GFMD and Eq. \eqref{eq:mean_gap}; $P_0(g, \zeta)$ predicted by GFMD, present work with and without corrected $\langle g \rangle$, and \texttt{Joe_etal_2023} under (b) $\bar{p}/E^* = 1 \times 10^{-4}$, (c) $\bar{p}/E^* = 1 \times 10^{-3}$, (d) $\bar{p}/E^* = 1 \times 10^{-2}$.}\label{fig:Fig_7}
\end{figure}

The PDF $P_0(g, \zeta)$ predicted by the present work shows a noticeable difference from the GFMD results under light load conditions, as shown in Figs. \ref{fig:Fig_4}(a-c). We revisit the same contact problem with three different normal loads, namely $\bar{p}/E^* = 1 \times 10^{-4}$, $1 \times 10^{-3}$ and $1 \times 10^{-2}$, to demonstrate the impact of correcting $\langle g \rangle$ on improving the accuracy of the predicted $P_0(g, \zeta)$. We generate a series of rough surface realizations numerically with a number of magnification values, i.e., $\zeta = 2, 4, 6, 8, 10, 20, 40, 60, 80, 100$. The spectra of all generated rough surfaces share the same phase angles. The magnification-dependent mean gaps under three light loads predicted by Eq. \eqref{eq:mean_gap} and GFMD are shown in Fig. \ref{fig:Fig_7}(a). We have already shown in Fig. \ref{fig:Fig_5}(b) that Eq. \eqref{eq:mean_gap} overestimates $\langle g \rangle$ at $\zeta = 100$. Fig. \ref{fig:Fig_7}(a) further confirms that this overestimation mainly occurs across almost the entire range of $\zeta \in [1, 100]$. Using the discrete data of $\langle g \rangle$ solved by GFMD, we can linearly interpolate $\langle g \rangle$ at any value of $\zeta \in [1, 100]$. The present work with the corrected $\langle g \rangle$ has a significant improvement of $P_0(g, \zeta)$, see Figs. \ref{fig:Fig_7}(b-d). When $\bar{p}/E^* = 1 \times 10^{-4}$, the improved $P_0(g, \zeta)$ (red line in Fig. \ref{fig:Fig_7}(b)) is globally shifted from the original $P_0(g, \zeta)$ (red solid dashed line in Fig. \ref{fig:Fig_7}(b)) toward lower gap ranges, and are almost the same as that predicted by \texttt{Joe_etal_2023}. When compared with GFMD, the improved present work slightly underestimates at low and overestimates at high gap ranges. Figs. \ref{fig:Fig_7}(c,d) show a similar improvement at higher loads ($\bar{p}/E^* = 1 \times 10^{-3}$ and $1 \times 10^{-2}$), particularly in the very small gap ranges where the improved model accurately reproduces the local peak. Comparing Fig. \ref{fig:Fig_7}(b-d) with Fig. \ref{fig:Fig_4}(a-c), the present work with corrected $\langle g \rangle$ demonstrates higher accuracy than two analytical solutions (\texttt{Almqvist_etal_11} and \texttt{Afferrante_etal_18}) across almost all the investigated gap ranges. Even though the computational cost of the present work is relatively higher than that of analytical solutions, Fig. \ref{fig:Fig_7}(b-d) implies that the diffusion-convection equation (Eq. \eqref{eq:convec_diff_equation}) can accurately describe the evolution of $P_0(g, \zeta)$ with magnification as long as an accurate mean gap is provided. 

Besides improving the prediction of the mean gap (drift coefficient), as we found above, the accuracy of the diffusion coefficient also influences the predicted $P_0(g, \zeta)$. We have used the magnification-dependent diffusion coefficient which were interpolated from those obtained by GFMD in the present work. We found that the agreement between the GFMD and the present work has deteriorated. Inspired by the superior accuracy of \texttt{Joe_etal_23}, as shown in Figs. \ref{fig:Fig_4} and \ref{fig:Fig_7}, the deteriorated results are strong signals that the approximated diffusion coefficient (Eq. \eqref{eq:B2_1}) should be replaced by its original form (Eq. \eqref{E:Bn_forward_KM_expansion}) to further improve its accuracy. Unlike the diffusion coefficient, $B_2(\zeta)$, in the present work that solely relies on magnification, the diffusion coefficient $B_2(g, \zeta)$ in \texttt{Joe_etal_23} is dependent on both gap and magnification. Therefore, future work should focus on the analytical or phenomenological form of the diffusion coefficient.

\begin{figure}[h!]
  \centering
  %\internallinenumbers 
  \includegraphics[width=9cm]{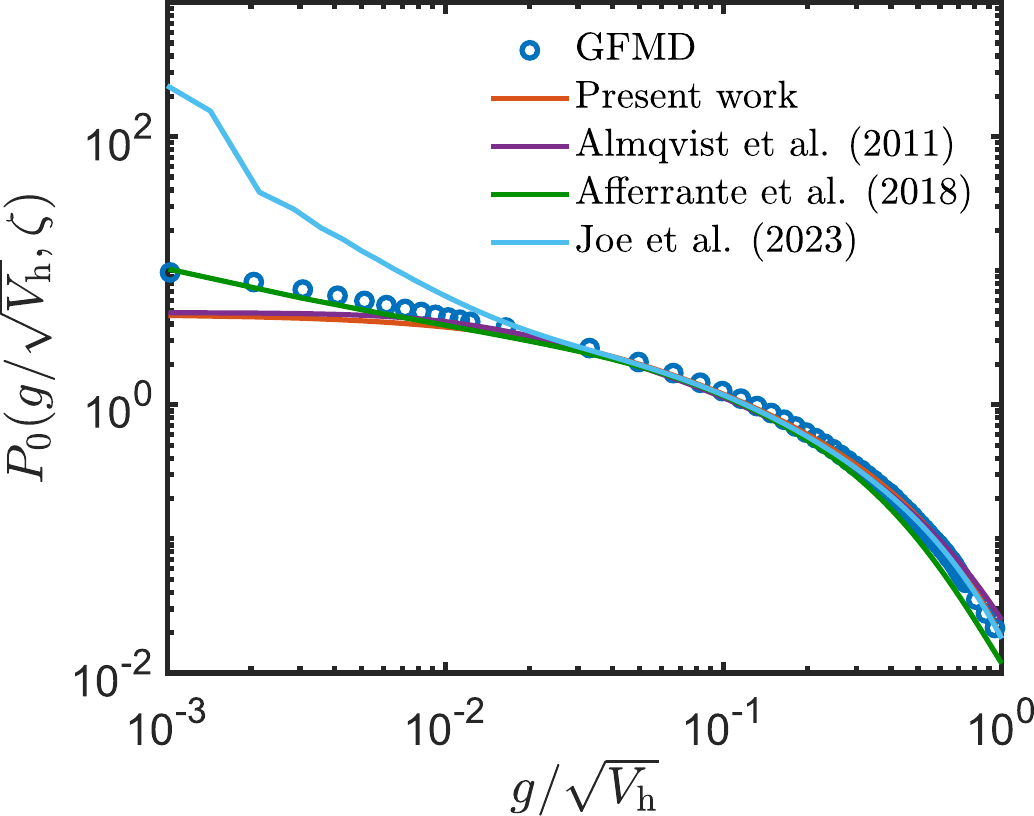}
  \caption{$P_0(g/\sqrt{V_{\text{h}}}, \zeta)$ predicted by various models with $\bar{p}/E^* = 0.086$. Relative contact area is $A^* = 0.5261$ (GFMD). The uncorrected form of $\langle g \rangle$ is used in the present work.}\label{fig:Fig_6}
\end{figure}

Fig. \ref{fig:Fig_4}(d) cannot illustrate the agreement between divergent trends of $P_0(g, \zeta)$ predicted by various models when the gap is extremely small. Fig. \ref{fig:Fig_6} illustrates $P_0(g, \zeta)$ in the log magnification, where $\bar{p}/E^* = 0.086$ and more than half of the interface is in contact. All theoretical models show good agreement with GFMD at higher gap ranges, and predictions start to bifurcate at $g/\sqrt{V_{\text{h}}} = 2 \times 10^{-2}$. The PDF curve predicted by \texttt{Afferrante_etal_18} has the best agreement with that by GFMD at $g/\sqrt{V_{\text{h}}} \leq 2 \times 10^{-2}$. The present work and \texttt{Almqvist_etal_11} have almost the same predictions, which are slightly lower than GFMD results. Predictions by \texttt{Joe_etal_23} are larger than GFMD results. As $g/\sqrt{V_{\text{h}}}$ decreases, the prediction by \texttt{Joe_etal_23} deviates more from GFMD results, with a maximum ratio of $20$ times at $g/\sqrt{V_{\text{h}}} = 1 \times 10^{-3}$. The reason for this overestimation at low interfacial gap ranges is because $P_0(g, \zeta)$, as predicted by \texttt{Joe_etal_23}, is the PDF of the interfacial gap over the entire interface, i.e., $\int_0^{\infty} P_0(g, \zeta) \text{d}g = 1$. Since there is no sharp distinction between the contact and non-contact areas in \texttt{Joe_etal_23}, the interfacial gap within the real contact area is extremely small (close to the atomic equilibrium distance), but not vanishing. Therefore, the non-vanishing interfacial gap within the real contact area, which is dominant over the entire interface, significantly increases the corresponding PDF at lower gap ranges. 

The no re-entry assumption used in this study, as well as in Persson's theory \cite{Persson01} and BEM by Bemporad and Paggi \cite{bemporad2015optimization}, is not strictly held. For example, in a rough surface contact simulation with a $4,096 \times 4,096$ sampling points, Dapp et al. \cite{Dapp14} found from the numerical results of GFMD that about 0.597$\%$ of the nominal contact area have ``re-entered" from out-of-contact status (at $\zeta = 128$) to contact status (at $\zeta = 129.5$). With a large increase in magnification (from $\zeta = 3.2$ to $\zeta = 32$), Dapp et al. \cite{Dapp14} showed that nearly $50 \%$ of the nominal contact area has experienced ``re-entry". The violation of the no re-entry assumption can lead to an overestimation of $P_0(g, \zeta)$ only when the local gap is small. This is because regions with smaller gap are more likely to transition to a contact status at higher magnifications. As a matter of fact, the present work may overestimate $\bar{A}^*$ (underestimate $A^*$) and the change of $\bar{A}^*$ relative to an increment of magnification. The violation's impact on the predicted $P_0(g, \zeta)$ accumulates during the diffusion-convection process, particularly for rough surfaces with a wide range in the wavenumber domain of their PSD. At low load condition (see Fig. \ref{fig:Fig_7}(b,c)), the present work, with and without corrected mean gap, underestimates $P_0(g, \zeta)$ at low gap ranges when compared with GFMD results. With high normal load (see Fig. \ref{fig:Fig_7}(d)), the present work with corrected mean gap overestimates $P_0(g, \zeta)$ at low gap ranges. This comparison may suggest that violating the no re-entry assumption may affect the stochastic process model under high normal loads, and is a secondary factor that influences its accuracy under low loads. This hypothesis can be supported by the fact that the interfacial gap decreases with the increasing normal load. Thus, points inside the non-contact area are more likely to re-enter the contact area under high loads as magnification increases. Future studies should focus on quantifying the error caused by violating the no re-entry assumption and abandoning this assumption in new stochastic process models.

\begin{figure}
  \centering
  \includegraphics[width=16cm]{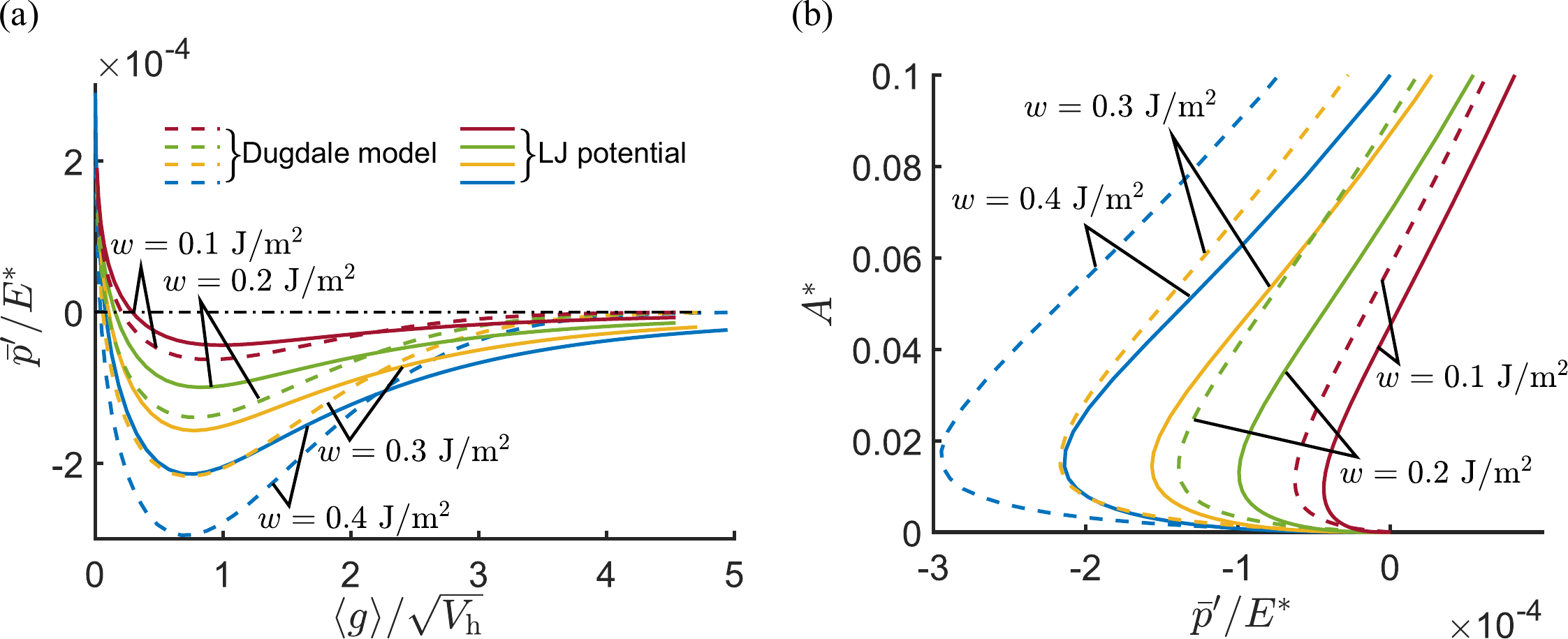}
  \caption{(a) Variations of $\bar{p}'/E^*$ with $\langle g \rangle/\sqrt{V_{\text{h}}}$. (b) Variations of $A^*$ with $\bar{p}'/E^*$. The PSD of rough surface is characterized by $q_{\text{l}} = q_{\text{r}} = 1 \times 10^5$ m$^{-1}$, $\zeta = 100$, $H = 0.8$, $\sqrt{V_{\text{h}}} = 0.6$ nm. Other essential parameters: $w$ = 0.1, 0.2, 0.3, 0.4 J/m$^2$, $\epsilon = 1$ nm, $E = 1 \times 10^{12}$ Pa, $\nu = 0.5$. The uncorrected form of $\langle g \rangle$ is used.}\label{fig:Fig_R7}
\end{figure}

One direct application of the present model is to solve the adhesion of rough surfaces under the Derjaguin-Muller-Toporov (DMT) limit \cite{ciavarella2018very}, i.e., $P_0(g, \zeta)$ and $P_0(p, \zeta)$ are independent of the surface interaction outside the contact area. The mean adhesive traction $\bar{p}' = \bar{p} + \bar{p}_{\text{ad}}$ where $\bar{p}_{\text{ad}}$ is the mean tensile traction due to surface interaction at non-contact area. Assuming the surface interaction is dominated by intermolecular interactions characterized by the Lennard-Jones (LJ) potential, 
\begin{equation}
\sigma_{\text{LJ}}(g) = \frac{8 w}{3 \epsilon}\left[ \left(\frac{\epsilon}{g + \epsilon}\right)^9 - \left( \frac{\epsilon}{g + \epsilon}\right)^3 \right],
\end{equation}
where $w$ is the work of adhesion and $\epsilon$ is the equilibrium distance ($\sigma_{\text{LJ}}(g = 0) = 0$), $\bar{p}_{\text{ad}}$ can be formulated directly as \cite{persson2014theory}
\begin{equation}\label{eq:p_ad_1}
\bar{p}_{\text{ad}} = \int_0^{\infty} P_0(g, \zeta) \sigma_{\text{LJ}}(g) \text{d}g.
\end{equation}
An alternative approach is to follow the bearing area model (BAM) proposed by Ciavarella \cite{ciavarella2018very} where surface interaction is simplified to a uniform distribution of $\sigma_{\text{th}} = -w/\epsilon$ over the bearing area with $g \in [0, \epsilon]$. This simplified interaction law is also known as the Dugdale model. Eq. \eqref{eq:p_ad_1} with $\sigma_{\text{LJ}}(g)$ approximated by the Dugdale model deduces to 
\begin{equation}\label{eq:p_ad_2}
\bar{p}_{\text{ad}} = \sigma_{\text{th}} \int_0^{\epsilon} P_0(g, \zeta) \text{d}g.
\end{equation}
In the original BAM \cite{ciavarella2018very}, the bearing area associated with $g \in [0, \epsilon]$ is approximated by truncating the cumulative distribution function of a Gaussian rough surface where the surface deformation outside the contact area is neglected. The present work can provide the same bearing area with better accuracy. Here, we apply the present work to predict the adhesive traction between a nominally flat surface (with the root mean square roughness in atomic scale) and a rigid flat. Fig. \ref{fig:Fig_R7}(a, b) shows respectively the predicted non-monotonic variations of $\bar{p}'$ with $\langle g \rangle$ and $A^*$ where $\bar{p}'$ first reaches the maximum tensile stress and followed by a monotonic increase in the compressive direction. These results are qualitatively the same as those predicted by BAM \cite{ciavarella2018very} and Persson's theory \cite{persson2014theory}. The $\bar{p}_{\text{ad}}$ predicted by Eq. \eqref{eq:p_ad_1} is slightly lower than that by Eq. \eqref{eq:p_ad_2}. 

In the current study, the evolutions of the PDF of the random contact pressure and random interfacial gap in contact and non-contact regions have been formulated consistently using two partial differential equations, diffusion and convection--diffusion equations, respectively, which belong to a broader concept of the Fokker--Planck equation, which makes the present work a unified framework for solving the PDFs of contact pressure and interfacial gaps based on   stochastic process theory, which we expect can be applied to solve the statistical behaviors of two complimentary stochastic variables in other mixed boundary value problems,  such as potential and current density on the electric contact interface, and interfacial shear stress and relative tangential surface displacement in the static frictional contact. 

\section{Conclusion}
We derived a convection--diffusion equation to describe the evolution of the PDF of an interfacial gap with the magnification. A finite difference method with unconditional stability was proposed to solve the convection--diffusion equation with nonnegative solutions. The predicted PDF of the interfacial gap   shows good agreement with that by GFMD, two variants of Persson's theory of contact, and the work of Joe et al., especially at high normal load ranges. Assuming that the variation of the random interfacial gap with the magnification is a Markov process, the obvious deviation associated with low load ranges is largely due to the overestimation of the mean interfacial gap and the oversimplified diffusion coefficient with no gap-dependency. We show in a typical problem that the accuracy of the present work is greatly improved at low load ranges if the mean interfacial gap solved by GFMD is used. Future studies should focus on improving the accuracy of the mean interfacial gap at low load ranges and formulating the gap- and magnification-dependent diffusion coefficient either analytically or phenomenologically. As one of its direct application, we show that the present work can effectively solve the adhesive contact problem under the DMT limit. The current study provides an alternative methodology for determining the PDF of the interfacial gap  and sheds light on a unified framework for solving the statistical behaviors of two complimentary stochastic variables in a mixed boundary value problem based on stochastic process theory.

\section{Acknowledgements}
This work was supported by the National Natural Science Foundation of China (No. 52105179), Fundamental Research Funds for the Central Universities (No. JZ2023HGTB0252), Natural Science Foundation of Jiangsu Province (No. BK20220555), and Anhui Provincial Natural Science Foundation (No. 2208085MA17). YX would like to thank Dr. Andrei Shvarts (University of Glasgow) for his helpful comments on the finite difference method. The authors would like to thank two anonymous reviewers for their insightful comments.  
\begin{spacing}{1}
	\bibliographystyle{gbt7714-nsfc}
	\bibliography{ref}
\end{spacing}
\appendix
\setcounter{figure}{0}
\setcounter{equation}{0}
\renewcommand{\thefigure}{A.\arabic{figure}}
\renewcommand{\theequation}{A.\arabic{equation}}
\renewcommand{\thetable}{A.\arabic{table}}

\section*{Appendix A. Persson's theory of contact for $P_0(g, \zeta)$}
Persson and his colleagues developed two variants of Persson's theory for $P_0(g, \zeta)$, an early version   \cite{Almqvist11}, and a newer version   \cite{Afferrante18} that is slightly different. Unfortunately, the expressions of the same theory given in two references \cite{Almqvist11,Afferrante18} are inconsistent. To fix this, we rewrite the core expressions of Persson's theory of contact for $P_0(g, \zeta)$ with minor changes. 

The PDF of the interfacial gap is written as a Riemann--Stieltjes integral, 
%\begin{linenomath*}
\begin{equation}\label{eq:Persson_PDF_gap}
P_0(g, \zeta) = -\int_1^{\zeta} \frac{1}{\sqrt{2 \pi h_{\text{rms}}^2(\zeta_1)}} \left[ \text{exp}\left(-\frac{(g - g_1(\zeta_1))^2}{2 h_{\text{rms}}^2(\zeta_1)} \right) + \text{exp} \left( -\frac{(g + g_1(\zeta_1))^2}{2 h_{\text{rms}}^2(\zeta_1)} \right) \right] \text{d}A^*(\bar{p}, \zeta_1), 
\end{equation}
%\end{linenomath*}
where $A^*(\bar{p}, \zeta_1)$ is determined by Eq. \eqref{eq:Area}. The variable $h_{\text{rms}}^2(\zeta_1)$ represents the variance of the rough surface height, including the spectral components with the wavenumber $q \in [\zeta_1 q_{\text{l}}, \zeta q_{\text{l}}]$, 
%\begin{linenomath*}
\begin{equation}
h_{\text{rms}}^2(\zeta_1) = 2 \pi \int_{\zeta_1 q_{\text{l}}}^{\zeta q_{\text{l}}} q C(q) \text{d} q 
\end{equation}
\begin{equation}
g_1(\zeta_1) = \bar{g}(\zeta_1) + \bar{g}'(\zeta_1) A^*(\bar{p}, \zeta_1)/A^{*'}(\bar{p}, \zeta_1), 
\end{equation}
%\end{linenomath*}
%\begin{linenomath*}
where the prime mark denotes the derivative with respect to $\zeta_1$. The variable $\bar{g}(\zeta_1)$ represents the mean interfacial gap, 
\begin{equation}\label{eq:mean_gap_1}
\bar{g}(\zeta_1) = \frac{\pi E^*}{2} \int_{\zeta_1 q_{\text{l}}}^{\zeta q_{\text{l}}} q^2 C(q) I(\bar{p}, q) \text{d} q, 
\end{equation}
where the expression of $I(\bar{p}, q)$ can be found in Eq. \eqref{eq:I_form}. The only difference between Eqs. \eqref{eq:mean_gap} and \eqref{eq:mean_gap_1} is the integral limits.\footnote{A typographic error can be found in the integral limits in Eq. (9) in Ref. \cite{Afferrante18}.} The above formulations belong to the older version, i.e., \texttt{Almqvist_etal_11}. A new version (\texttt{Afferrante_etal_18}) replaces $h_{\text{rms}}$ in Eq. \eqref{eq:Persson_PDF_gap} with 
\begin{equation}
h_{\text{rms}}^{\text{eff}}(\zeta_1) = \left[h_{\text{rms}}^{-2}(\zeta_1) + g_1^{-2}(\zeta_1)\right]^{-1/2}. 
\end{equation}
%\end{linenomath*}
\end{document}